\documentclass{acm_proc_article-sp}
\usepackage{graphicx}
\newtheorem{lem} {Lemma} 
\newcommand {\BL} {\begin{lem}} 
\newcommand {\EL} {\end{lem}} 

\newtheorem{cor}  {Corollary} 
\newcommand {\BCR} {\begin{cor}}
\newcommand {\ECR} {\end{cor}}

\newtheorem{thm} {Theorem} 
\newcommand {\BT} {\begin{thm}}
\newcommand {\ET} {\end{thm}}

\newtheorem{defi} {Problem} 

\newcommand {\BDE} {\begin{defi}}
\newcommand {\EDE} {\end{defi}}
\newcommand{\comment}[1]{}

\newtheorem{xxx} {Definition} 
\newcommand {\BD} {\begin{xxx}}
\newcommand {\ED} {\end{xxx}}

\newtheorem{xxxcc} {Problem} 
\newcommand {\BP} {\begin{xxxcc}}
\newcommand {\EP} {\end{xxxcc}}

\newtheorem{xxxxx} {Observation} 
\newcommand {\BO} {\begin{xxxxx}}
\newcommand {\EO} {\end{xxxxx}}

\begin{document}

\title{On Optimal Top-{\huge$\boldsymbol{k}$} String Retrieval}
%
%
%
%
%

\numberofauthors{4} 
%
\author{
%
%
\alignauthor
Rahul Shah\\
       \affaddr{ Louisiana State University, USA.}\\
       \email{rahul@csc.lsu.edu}
\alignauthor
Cheng Sheng\\
       \affaddr{The Chinese University of Hong Kong, China.}\\
       \email{csheng@cse.cuhk.edu.hk}
\alignauthor Sharma V. Thankachan\\
         \affaddr{ Louisiana State University, USA.}\\
       \email{thanks@csc.lsu.edu}
\and  
\alignauthor Jeffrey Scott Vitter \\
       \affaddr{The University of Kansas, USA.}\\
       \email{jsv@ku.edu}
}


\maketitle
\begin{abstract}

Let ${\cal{D}}$ = $\{d_1, d_2, d_3, ..., d_D\}$ be a given set of $D$ (string) documents of total length $n$. The top-$k$ document retrieval problem is to index $\cal{D}$ such that when a pattern $P$ of length $p$, and a parameter $k$ come as a query, the index returns the $k$ most relevant documents to the pattern $P$. Hon et. al.~\cite{HSV09} gave the first linear space framework to solve this problem in $O(p + k\log k)$ time. This was improved by Navarro and Nekrich~\cite{NN12} to $O(p + k)$. These results 
are powerful enough to support arbitrary relevance functions like frequency, proximity, PageRank, etc. In many applications like desktop or email search, the data resides on disk and hence disk-bound indexes are needed. Despite of continued progress on this
problem in terms of theoretical, practical and compression aspects, any non-trivial bounds in external memory model have so far been elusive. 
Internal memory (or RAM) solution to this problem decomposes the problem into $O(p)$ subproblems and thus incurs the additive factor of $O(p)$. In external memory, these approaches will lead to $O(p)$ I/Os instead of optimal $O(p/B)$ I/O term where $B$ is the block-size. We re-interpret the problem independent of $p$, as interval stabbing with priority over tree-shaped structure. This leads us to a linear space index  in external memory supporting top-$k$ queries (with unsorted outputs) in
near optimal $O(p/B + \log_B n + \log^{(h)} n + k/B)$ I/Os for any constant $h$\footnote{$\log^{(1)}n =\log n$ and $\log^{(h)} n = \log (\log^{(h-1)} n)$}. Then we get $O(n\log^*n)$ space index with optimal $O(p/B+\log_B n + k/B)$ I/Os.
As a corollary, we also show the result in RAM which allows sorted order retrieval in $O(k)$ time, if the locus of pattern match is provided in advance. This gives optimal performance in many applications where finding the locus of pattern in a suffix tree can be done much faster than usual $O(p)$.

\end{abstract}

%
 \terms{Theory}

\section{Introduction}

The inverted index is the most fundamental data structure in the field of information retrieval~\cite{ZM06}.
It is the backbone of every known search engine today.
For each word in any document collection,
  the inverted index maintains a list of all documents in that collection which
  contain the word. 
Despite its power to answer various types of queries,
  the inverted index becomes inefficient,
  for example, when queries are phrases instead of words.
This inefficiency results from inadequate use of
  word orderings in query phrases~\cite{PTS+11}.
Similar problems also occur in applications
  when word boundaries do not exist or cannot be identified deterministically
  in the documents, like genome sequences in bioinformatics
  and text in many East-Asian languages.
These applications call for data structures to answer queries in
  a more general form, that is, (string) pattern matching.
Specifically, they demand the ability to
  identify efficiently all the documents that
  contain a specific pattern as a substring.
The usual inverted-index approach might require
  the maintenance of document lists for
  all possible substrings of the documents.
This can take quadratic space and
  hence is neither theoretically interesting
  nor sensible from a practical viewpoint.

The first frameworks for answering pattern matching (and related) queries
  were proposed by Matias et.~al.~\cite{MMSZ98} and Muthukrishnan~\cite{Muth02}.
Their data structures solve the \emph{document listing problem},
  in which a collection $\mathcal{D}$
  of $D$ documents is required to be indexed
  so that given a query pattern $P$ of length $p$,
  all the documents that contain $P$ can be retrieved efficiently.
As the pattern can appear in a single document multiple times,
  a major challenge of this problem is that
  the overall number of the pattern occurrences can be much greater
  than the number $ndoc$ of the result documents.
Therefore, it is unaffordable to answer a query by
  enumerating all the occurrences of $P$.

Muthukrishnan also initiated the study of
  relevance metric-based document retrieval~\cite{Muth02},
  which was then formalized by Hon et al.~\cite{HSV09}
  as the \emph{top-$k$ document retrieval problem}.
Here,
  instead of all the documents that match a query pattern,
  the problem is to output the $k$ documents most relevant to the query in sorted order of relevance score.
Relevance metrics considered in the problem can be
  either pattern-independent (eg., PageRank)
  or -dependent.
In the latter case one can take into account information like
  the frequency of the pattern occurrences 
(or term-frequency of popular tf-idf measure)
  and even the locations of the occurrences (e.g.,\emph{min-dist}~\cite{HSV09} which takes  proximity of two closest occurrences of pattern as the score).
The framework of Hon et al.~\cite{HSV09} takes linear space
  and answers the query in $O(p + k\log k)$ time.
This was then improved by Navarro and Nekrich~\cite{NN12} to achieve $O(p+ k)$ query cost,
  which is in a way optimal. Several other approaches for top-$k$ document retrieval have recently been published.
Some use, instead of linear space, succinct space~\cite{bell,HSV09}
  or semi-succinct space~\cite{valimaki,puglisi,navarrotalk,sharmacpm,bell}.
Their query costs, however,  usually contain a multiplicative poly-logarithmic factor
  to the output size $k$ (or $ndoc$). This problem is seeing a burst of research activity in both mainstream venues for algorithms and information retrieval ~\cite{HSV09,soda11,NN12,PTS+11} as well as plenary talks~\cite{navarrotalk,vittertalk} in the string matching community.

With the advent of enterprise search, deep desktop search, email search technologies, the indexes which reside on disks are more and more important. Many biological databases are now being turned for external memory versions as the amount of sequence data grows.
String retrieval is central to biological and image data as well as phrase querying in text files.
Even though there has been series of work on internal memory top-$k$ string, including in theory as well as practical IR~\cite{PTS+11,sigir12} communities, most implementations (as well as theoretical results) have focused on RAM based compressed and/or efficient indexes. Newer applications such as desktop music search or email search sometimes also have additional range constraints on different attributes like date of the email, or sender etc. For typical practical values, both $p/B$ and $k/B$ factors turn out to be some constant I/Os, and thus rightly designed I/O efficient index can make the difference between hundreds of I/Os vs a small constant number of I/Os. Despite these motivations as I/O-efficient index for string retrieval has been elusive. This is perhaps because the problem has been modeled as 
4-sided orthogonal range queries in 3-dimensions.
There is no hope for I/O-efficient range query structure for this more general problem.

Apart from an external memory index, there are also RAM situations where current internal memory bounds are inadequate.
All the internal memory approaches use a two-phase procedure to answer a query. The first phase identifies the \emph{locus} of $P$ in a suffix tree,  that is, the node corresponding to the pattern $P$.
The second phase finds the top-$k$ results in the subtree rooted at the locus.
\cite{HSV09} and \cite{NN12} reduce the Phase-2 subproblem to
  a 4-sided orthogonal range searching problem in 3D.
While general four-sided orthogonal range searching is proved hard~\cite{Chaz90},
  the desired bounds can nevertheless be achieved by identifying a special property that
  one dimension of the reduced subproblem can only have $p$ distinct values.
Employing this property,
  an additive $O(p)$-term inevitably appears
  in the cost to handle Phase-2,
  which is actually sub-optimal.
Our technique does the Phase-2 independent of $p$. Thus, we show that
  Phase-2 can be answered strictly in $O(k)$ time in the RAM model.
There are many string matching applications where finding the locus can be done much faster than $O(p)$.
This is especially true if application works on finding multiple loci typically with constant time per locus.
In applications like cross-document pattern matching \cite{KNS12}, pattern $P$ is given by a location in some document and is needed to be found in other documents.
Since the collection can be pre-indexed, the locus of the pattern can be found in $O(\log\log n)$ using weighted level ancestor query (or even in faster $O(\log\log p)$ time).
In many pattern matching applications, for example in suffix-prefix overlap~\cite{cpm10} , maximal substring matches~\cite{bojian}, or autocompletion search (like in Google Instant$^{TM}$)
multiple loci are searched with amortized constant time for each locus.
In such situations, having extra $O(p)$ from the Phase-2 leads to inefficient solutions.


\subsection{Related Work, Problem Complexity and Our Results}

For the document listing problem,
  Muthukrishnan gave a somewhat optimal solution,
  which uses linear space and $O(p+ ndoc)$ query cost~\cite{Muth02}.
As the overall number of the occurrences of $P$ can be  much larger than $ndoc$,
  he uses the idea of \emph{chaining},
  by which a one-sided constraint on a particular dimension guarantees that
  no document can be enumerated more than once.
With proper labeling, 
  the pattern can be converted into a $2d$ orthogonal 3-sided query.

\BP
{\bf Top-$k$ document retrieval problem: }\\
Let $w(P,d)$ be the score function which depends on the set of occurrence locations of pattern $P$  in document $d$.
Given a document collection $\mathcal{D}$$= \{d_1, d_2, ..., d_D\}$ of $D$ documents, build an index answering the following query:
given $P$ and $k$, find $k$ documents  with the highest $w(P,d)$ values in its sorted (or unsorted) order.
\EP

For example, one of the popular score function is the frequency with which $P$ occurs in $d$.
In this case, we can imagine that each leaf of the suffix tree is annotated with the document-id to which that suffix belongs. Then, 
the problem boils down to listing top-$k$ most frequently appearing document ids in the suffix range of the pattern. Again, each document should appear 
exactly once.

Hon et al. extended the idea to tree-shaped chaining,
  which can be used to solve the top-$k$ document retrieval problem~\cite{HSV09}.
Here, the additional top-$k$ constraint can be converted to a threshold in the third dimension,
  thus resulting in a 4-sided query in $3d$ space.

Both \cite{HSV09} and \cite{NN12} (which eventually achieves $O(p+k)$ time) use the fact that on one of the dimensions,
  the set of the geometric points related to query $P$ have only $p$ distinct values.
However, this speciality of the 4th constraint (uniqueness) cannot be exploited in external memory model.
We surprisingly show that even though this is a query with 4 constraints, we can almost achieve results similar to 
3-constrained queries. 

A related but somewhat orthogonal line of research has been to get top-$k$ or threshold queries on general array based ranges. In this, we are given array $D$ of colors, and for a range query $(i,j)$, we have to output top-most colors in this range (with each color reported at most once). If the the scoring criteria is based on frequency, then lower-bounds on range-mode problem\cite{GreveJLT10,ChanDLMW12} would imply no efficient (linear space and polylog time) data structures can exist. There are variants considered where each entry in the array has a fixed score or each color (document) has a fixed score, independent of number of occurrences. Recent ~\cite{soda13} surprising result (similar to ours) has been for 3-sided categorical range reporting where each entry has another attribute called score, and the query specifies range as well as score threshold. We are supposed to output all colors whose at least one entry within the range satisfies the score criteria. There are easier variants
where each entry of the same color gets the same score attribute~\cite{nekrich} like PageRank which have been shown to have efficient external memory results.
For even simpler variants, where only top-$k$ scores are to be reported \cite{LarsenP12, AfshaniBZ11} without considering colors or where all the colors (documents) have the same score and each document in the range must be uniquely reported (as in \emph{document listing}). Both these variants lead to 3-sided queries which are easier to solve in external memory. However, this line of results, does not consider, set-based score functions like frequency or proximity. Partly parallel to our work, the result of \cite{soda13} though achieves surprising result of achieving optimal I/Os with $O(n\log^*n)$ space for 3-sided categorical range reporting. This work has a similar surprise factor as ours, in the sense that its traditional modeling would require 4-sided queries (with 4th constraint coming from the uniqueness of color reporting). 
However, this line of work or array ranges, due to its theoretical limitations, cannot support arbitrary score functions which we require here.


We summarize our results as follows.
\begin{enumerate}
\item
In RAM, there exists an $O(n)$-word data structure
  that solves top-$k$ (sorted) document retrieval problem 
  in $O(k)$ time, once the locus of the pattern match is given.
This result improves the previous work~\cite{HSV09,NN12} by eliminating the additive term $p$.
\item
In the external memory model, there exists an $O(nh)$-word structure
  that solves the top-$k$ (unsorted) \footnote{It is known that if we need to get sorted reporting in external memory, then $k/B$ term needs to be at least $((k/B)\log_{M/B} (k/B))$~\cite{AfshaniBZ11}} document retrieval problem
    in $O(p/B+ \log_B n + \log^{(h)} n + k/B)$ I/Os for any $h\leq \log^*n$.
  This implies that optimal $O(p/B+ \log_B n + k/B)$ query  I/Os can be achieved using an almost-linear $O(n\log^* n)$-word space structure.

\end{enumerate}

\section{Preliminary: Top-{\large \lowercase{$\boldsymbol{{k}}$}}  Framework}\label{sec:prelim}

This section briefly explains the framework of Hon et.~al.~\cite{HSV09}.
  The \emph{score} of a document $d$ with respect to a pattern $P$ denoted by  $w(P,d)$
  be the relevance of $d$ to $P$,
  which is a function of the locations of all $P$'s occurrences in $d$.
The generalized suffix tree (GST)
  of a document collection $\cal{D}$$ = \{d_1,d_2,d_3,\ldots,d_D\}$
  is the combined compact trie (a.k.a.\ Patricia trie)
  of all the non-empty suffices of all the documents.
Use $n$ to denote the total length of all the documents,
  which is also the number of the leaves in GST.
For each node $u$ in GST, consider the path from the root node to $u$.
Let $depth(u)$ be the number of nodes on the path,
  and $prefix(u)$ be the string obtained by
  concatenating all the edge labels of the path.
For a pattern $P$ that appears in at least one document,
  the \emph{locus} of $P$, denoted as $u_P$,
  is the node closest to the root satisfying that $P$ is a prefix of $prefix(u_P)$.
By numbering all the nodes in GST in the pre-order traversal manner,
  the part of GST relevant to $P$ (i.e., the subtree rooted at $u_P$)
  can be represented as a range.  
  
Nodes are marked with documents.
A leaf node $\ell$ is marked with a document $d \in \cal{D}$
  if the suffix represented by $\ell$ belongs to $d$.
An internal node $u$ is marked with $d$
  if it is the lowest common ancestor of two leaves marked with $d$.
Notice that a node can be marked with multiple documents.
For each node $u$ and each of its marked document $d$,
  define a \emph{link} to be a quadruple $(origin, target, doc, score)$,
  where $origin=u$,
  $target$ is the lowest proper ancestor%
  \footnote{Define a dummy node as the parent of the root node, marked with all the documents.}
  of $u$ marked with $d$,
  $doc=d$ and
  $score=w\bigl(prefix(u), d\bigr)$.
Two crucial properties  of the links identified in ~\cite{HSV09} are listed below.

\BL
\label{lem_linkunique}
For each document $d$ that contains a pattern $P$,
  there is a unique link whose origin is in the subtree of $u_P$
  and whose target is a proper ancestor of $u_P$.
The score of the link is exactly the score of $d$ with respect to $P$.
\EL

\BL
\label{lem_linkcount}
The total number of links is  $O(n)$. 
\EL

Based on Lemma~\ref{lem_linkunique}, the top-$k$ document retrieval problem can be reduced to
  the problem of finding the top-$k$ links (according to its score) stabbed by $u_P$, where \emph{link stabbing} is defined as follows: 
  
  \BD[Link Stabbing]
We say that a link  is \emph{stabbed} by node $u$ if it is originate in the subtree of $u$ and  target at a proper ancestor of $u$. 
\ED

If we order the nodes in GST as per the \emph{pre-order traversal} order,
  these constraints translate into finding all the links
  (i)~the numbers of whose origins fall in the number range of the subtree of $u_P$, and
  (ii)~the numbers of whose targets are less than the number of $u_P$.
Regarding constraint~(i) as a two-sided range constraint on x-dimension,
  and regarding  constraint~(ii) as a one-sided range constraint on y-dimension,
  the problem asks for the top-$k$ weighted points  that fall in a three-sided window in 2d space, where weight of a point be the score of the corresponding link.

\section{External Memory Structures}
This section is dedicated to describing our external memory data structures.
The initial phase of pattern search can be performed in $O(p/B+\log_B n)$ I/O's using a string B-tree. Once the suffix range of $P$ is identified, 
we take the lowest common ancestor of the left-most and right-most leaves in the suffix range of GST to identify the locus node $u_P$. 
Hence, the first phase (i.e., finding the locus node $u_P$ of $P$) takes optimal I/O's and now we focus only on the second phase (which we call as the retrieval phase). Our main result is summarized in the following theorem. 

\BT
\label{the_result01}
In the external memory model, there exists a $O(nh)$-word structure
  that solves the top-$k$ (unsorted)  document retrieval problem
  in $O(p/B+ \log_B n + \log^{(h)} n + k/B)$ I/Os for any $h \leq \log^*n$. 
\ET

\BCR
 Top-$k$ (unsorted)  document retrieval problem in external memory can be solved in optimal $O(p/B+ \log_B n + k/B)$ query  I/Os using an  $O(n\log^* n)$ word space data structure. 
\ECR

Instead of solving the top-$k$ version, we first solve a threshold version in Sec~\ref{bdsp} where the objective is to retrieve those links stabbed by the query node $u_P$  with $score$ at least a given threshold $\tau$. 
Then in Sec~\ref{conversion}, we propose a separate structure that converts the original top-$k$-form query into a threshold-form query so that the the structure in Sec~\ref{bdsp}
can now be used to answer the original problem. 
Finally, we obtain Theorem~\ref{the_result01} via bootstrapping on a special structure for handling top-$k$ queries in lesser number of I/Os for small values of $k$.

\subsection{Breaking Down Into Sub-Problems} \label{bdsp}

We decompose the original problem into simpler sub-problems. 
Instead of solving the top-$k$ version, we first solve a threshold version, where the objective is to retrieve those links stabbed by $u_P$ with $score$ at least a given threshold $\tau$. 
The origin, target and score of a link $L_i$ are represented by $o_i, t_i$ and $w_i$ respectively. 

\BL \label{threshold}
There exists an $O(n)$ space data structure for answering the following query: given a query node $u_P$ and a threshold $\tau$, all links stabbed by $u_P$ with score $\geq \tau$ can be reported in $O(\log(n/B)+z/B)$ I/Os, where $z$ is the number of outputs. 
\EL

\paragraph{Rank and Components}
For any node $u$ in GST, we use $u$ to denote its pre-order rank as well. Let $size(u)$ denotes the number of leaves in the subtree of $u$,  then we define its  \emph{rank} as:
 $$rank(u) = \lfloor \log \lceil \frac{size(u)}{B} \rceil \rfloor $$

Note that $rank(.) \in [0, \lfloor \log \lceil \frac{n}{B} \rceil \rfloor] $. 
A contiguous subtree consisting of nodes with the same rank is defined as a \emph{component}, and the $rank$ of a component is same as the rank of nodes within it (see figure 1). 
Therefore, a \emph{component}  with $rank =0$ is a bottom level subtree of size (number of leaves)  at most $B$.
From the definition, it can be seen that a node and at most one of its child can have the same $rank$. 
Therefore, a  component with $rank = \delta \geq 1$  consists of nodes in a path which goes top-down in the tree. 

 \begin{figure}
\begin{center}
\includegraphics[scale=0.44]{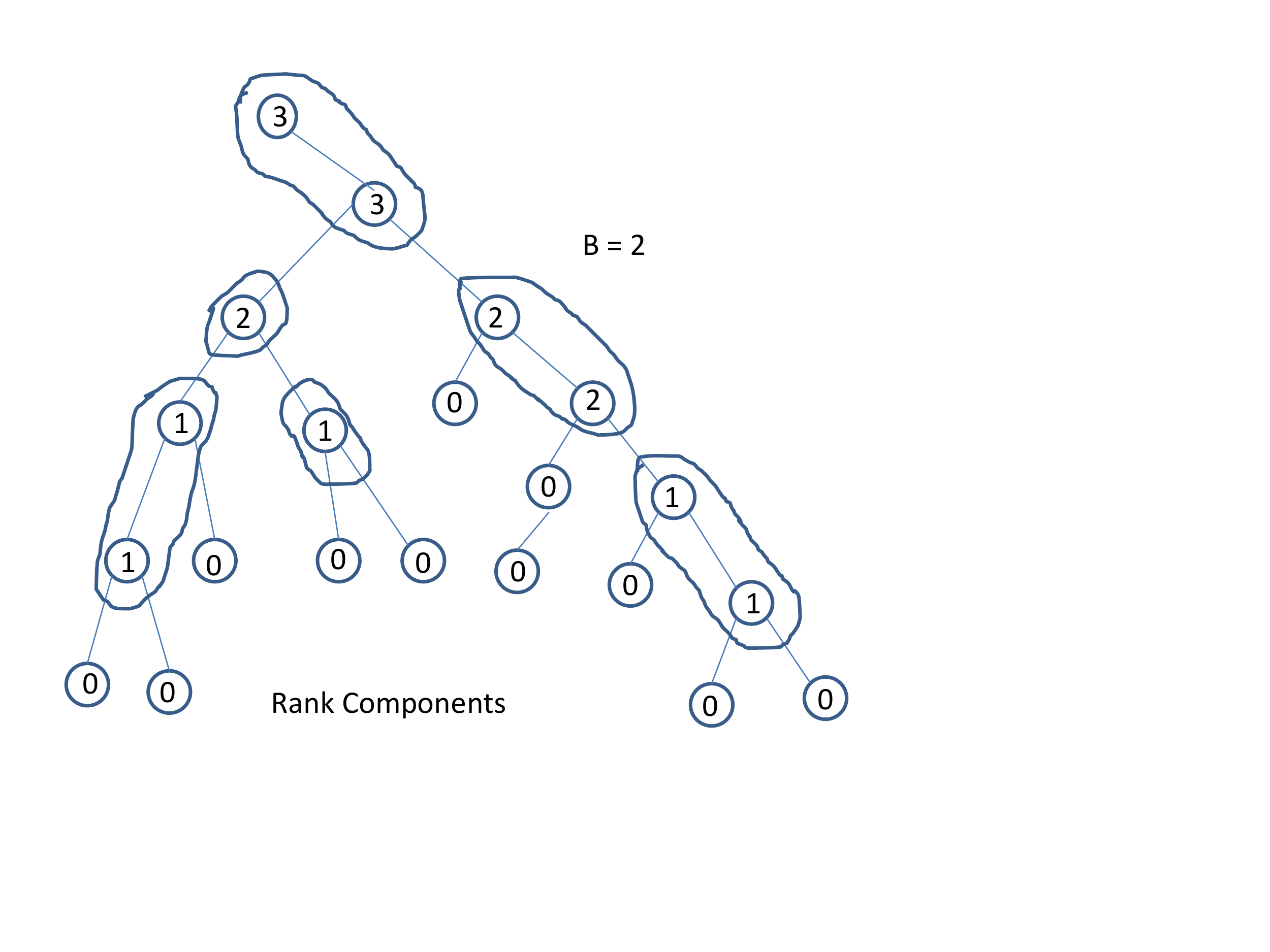}
\end{center}
\caption{Rank Components}
\end{figure}

The number of links originating within the subtree of any node $u$ is at most $2size(u)$. 
Therefore, the number of links originating within  a component with rank$=0$ is at most $2B$. These $O(B)$ links corresponding to each component with $rank=0$ can be maintained separately as a list, taking total $O(n)$ words space. Now, given a locus node $u_P$, if $rank(u_P) =0$, the number of links originating within the subtree of $u_P$ is also $O(B)$ and all of them can be processed in $O(1)$ I/O's by simply scanning the list of links corresponding to the component to which $u_P$ belongs to.

The query processing is more sophisticated when $rank(u_P) \geq 1$. For handling this case, 
we define the $rank$ for a link as the $rank$ of its target. Let $L_i$ be a link with $o_i, t_i$ and $w_i$  representing its  origin, target and score respectively, then: 
 $$rank(L_i) = rank(t_i)$$

We classify the links into the following 3 types based on its $rank$ with respect to the $rank$ of query node $u_P$: \\
\\ \emph{low-ranked links}: links with $rank(target) < rank(u_P)$
\\ \emph{equi-ranked links}: links with $rank(target) = rank(u_P)$
\\ \emph{high-ranked links}: links with $rank(target) > rank(u_P)$

None of the low-ranked links can be an output as their target will not be an ancestor of $u_P$, hence all low-ranked links can be ignored while querying. 
Therefore, we need to check for outputs among only equi-ranked and high-ranked links. 
Next in Sec~\ref{equi-pro} we show that the problem of retrieving outputs among equi-ranked links can be reduced to a 3d dominance query.
Then in Sec~\ref{high-pro} we show that the problem of retrieving outputs among high-ranked links can be reduced to at most $\lfloor \log \lceil \frac{n}{B} \rceil \rfloor$ 3-sided range queries in $2d$.

\subsubsection{Processing equi-ranked links} \label{equi-pro}

Let $C$ be a component and $S_C$ be set of all links $L_i$,  such that its target $t_i$ is a node in $C$. 
Also, for any link $L_i \in S_C$, let pseudo$\_$origin $s_i$ be the lowest ancestor of its origin $o_i$ within $C$ (see Figure 2).
Then a link $L_i \in S_C $  originates in the subtree of any node $u$ within $C$ if and only if $s_i \geq u$. 
Now if the locus $u_P$ is a node in $C$, then among all equi-ranked links, we need to consider only those links $L_i \in S_C$, because for any other equi-ranked link $L_j \notin  S_C$,  $o_j$ will not be in the subtree of $u_P$.  Based on the above observations, all equi-ranked output links  are those $L_i 
\in S_C$ with $t_i < u_P \leq s_i$ and $w_i \geq \tau$. By considering each link as a weighted interval with $t_i, s_i$ and $w_i$ represents the starting point, ending point and weight respectively, the above query can be reduced to the following:

 \begin{figure}
\begin{center}
\includegraphics[scale=0.44]{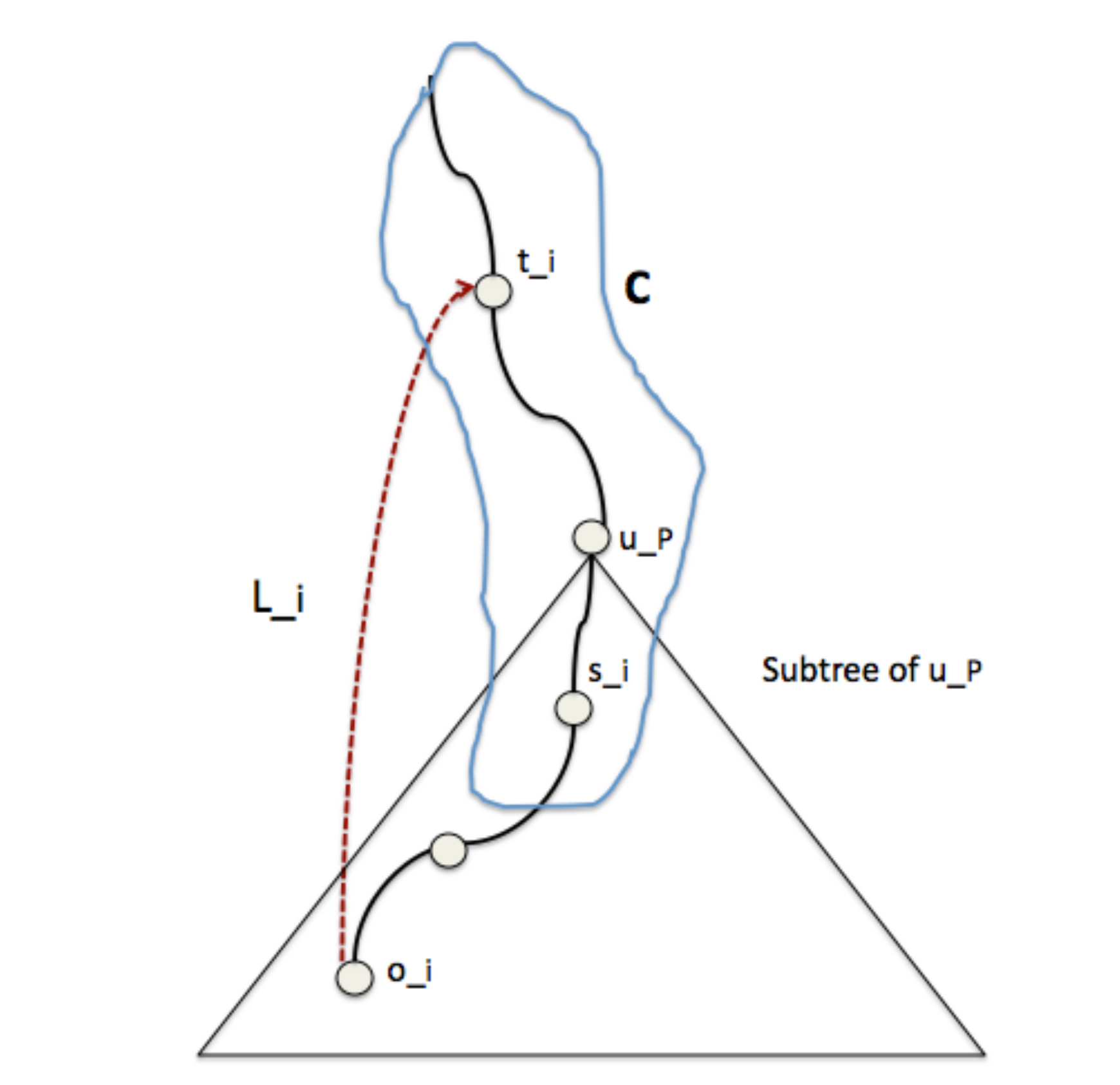}
\end{center}
\caption{Pseudo Origin}
\end{figure}

\BP
{\bf Interval stabbing with priority:}
Let $\cal{I}$ be a set of weighted intervals, build an index for answering the following query: given a query point $x$ and a threshold $\tau$, report all the intervals  stabbed by $x$ with weight at least  $\tau$.
\EP
Later we show how to solve this problem efficiently in internal memory.
To solve this in external memory, we 
 treat each link $L_i \in S_C$ as a 3d point $(t_i,s_i,w_i)$ and maintain a 3d dominance query structure over it. 
 Now the outputs with respect to $u_P$ and $\tau$ are those links corresponding to the points within $ [-\infty, u_P) \times [ u_P, \infty] \times [\tau, \infty]$. 
Such a structure for $S_C$ can be maintained in linear $O(|S_C|)$ words space and can answer the query in $O(\log_B |S_C|+z_{eq}/B)$ I/O's using the result by Afshani~\cite{Afshani08}, where $|S_C|$ is the number of points (corresponding to links in $S_C$) and $z_{eq}$ be the output size. Thus the over all space for maintaing these structures for all components is $O(n)$ words.

\BL
\label{lem_equidensity}
Given a query node $u_P$ and a threshold $\tau$, the links  stabbed by $u_P$ with score $ \geq \tau$ and rank equal to $rank(u_P)$ can be retrieved in 
$O(\log_B n+z_{eq}/B)$ I/Os using an $O(n)$ word space data structure, where $z_{eq}$ is the output size.
\qed
\EL

\subsubsection{Processing high-ranked links}\label{high-pro}

The following is an important observation. 
\BO
Any link $L_i$ with its origin $o_i$ within the subtree of a node $u$ is  stabbed by $u$ if $rank(L_i) > rank(u)$. 
\EO

This implies, while looking for the outputs among the high-ranked links,  the condition of $t_i$  being a proper ancestor of $u_P$ can be ignored as it is taken care of  automatically if $o_i\in \bigl[u_P, u_P'\bigr]$,  where $u_P'$ be the (pre-order rank of) right-most leaf in the subtree rooted at $u_P$. 
Let $G_r$  be the set of all links with rank equals $r$ for $1\leq r \leq \lfloor \log \lceil \frac{n}{B} \rceil \rfloor$.
Since there are only $O(\log(n/B))$ sets, we shall maintain separate structures for links in each $G_r$ by considering only $origin$ and $score$ values. 
We treat each link $L_i \in G_r$ as a $2d$ point $(o_i, w_i)$, and maintain a 3-sided range query structure over them for $r=1,2,..., \lfloor \log \lceil \frac{n}{B} \rceil \rfloor$. 
All high-ranked output links can be obtained by retrieving those links in $L_i\in G_r$ with the corresponding point $(o_i, w_i) \in [u_P, u_P'] \times [\tau, \infty]$ for $r =  rank(u_P)+1, ..., \lfloor \log \lceil \frac{n}{B} \rceil \rfloor$. By using the linear space data structure in~\cite{ArgeSV99}, the space and I/O bounds for a particular $r$ is given by $O(|G_r|)$ words and $O(\log_B|G_r|+z_r/B)$, where $z_r$ is the number of output links in $G_r$. Since a link can be a part of at most one $G_r$, the total space consumption is $O(n)$ words and the total query I/Os is $O(\log_B n \log (n/B)+z_{hi}/B)$, where $z_{hi}$ represents the number of high-ranked output links.

Next, we show how to improve the I/O bound to $O(\log (n/B)+z_{hi}/B)$. 
For this, we use the following result by Larsen and Pagh~\cite{LarsenP12}: a set of $m$ points in $2d$ on an $[0, m]\times [0, m]$ grid can be maintained in $O(m)$ word space and can answer a
space 3-sided range query  in optimal $O(1+z/B)$ I/O's, where $z$ is the output size. 
Therefore, we first reduce the points ($origin$ and $score$ values) into rank-space and then maintain this structure over them. 
It remains to show how to map the query parameters
 $u_P, u_P'$ and $\tau$ into rank-space, which are to be fed to the data structure.

 Firstly, we show how to compute the values corresponding to  $\bigl[u_P, u_P' \bigr]$ in rank-space. 
 For $1\leq r\leq \lfloor \log \lceil \frac{n}{B} \rceil \rfloor$,
  define $OR_{r}$ to be the array of all the links in $G_r$ 
  ordered by their $origin$. Our task is to compute the maximal subrange $[a...b]$ in $OR_r$, such that $u_P\leq a \leq b \leq u_P'$. 
 This conversion can be supported using a multiset $B_r'$ which consists of the $origin$ values of all links in $OR_r$. 
 We can store $B_r'$  with the succinct dictionary of~\cite{RRS07} in $O(n)$ bits and can perform the above conversion in $O(1)$ time. 
 Therefore, the total space for storing all multisets corresponding $ r= 1,2, \ldots \leq \lfloor \log \lceil \frac{n}{B} \rceil \rfloor$ is $O(n)$ words.
Since we need to perform  $O(\log(n/B))$ such constant time (I/O's) queries corresponding to $r = rank(u_P)+1, ..., \lfloor \log \lceil \frac{n}{B} \rceil \rfloor$, we need additional $O(\log(n/B))$ I/O's. 

Rank-space reduction of threshold $\tau$ for  all values of $r$ can also be performed in $O(\log(n/B))$ I/O's using another $O(n)$ word space structure. 
For $1\leq r\leq \lfloor \log \lceil \frac{n}{B} \rceil \rfloor$,
define $TH_{r}$ to be the array of all the links in $G_r$,
sorted in the increasing order of their $scores$. 
Given a threshold $\tau$ and $r$, find the minimum $c$ such that $TH_r[c] \geq \tau$. 
  This conversion can be supported using a multiset $B''_r$ which consists of the $scores$ of all links in $TH_r$. 
 As before we can store $B''_r$  with the succinct dictionary of~\cite{RRS07} in $O(n)$ bits and can perform the above conversion in $O(1)$ time. 
 Therefore, in $O(n)$ words  all the multisets corresponding $ r= 1,2, \ldots \leq \lfloor \log \lceil \frac{n}{B} \rceil \rfloor$ can be maintained and perform these conversions in total $O(\log(n/B))$ I/O's.

\BL
\label{lem_overdensity}

Given a query node $u_P$ and a threshold $\tau$, the links  stabbed by $u_P$ with score $ \geq \tau$ and rank $ \geq rank(u_P)$ can be retrieved in 
$O(\log_B n+z_{hi}/B)$ I/Os using an $O(n)$ word space data structure, where $z_{hi}$ is the output size.
\qed
\EL

By combining Lemma~\ref{lem_equidensity} and Lemma~\ref{lem_overdensity}, we obtain Lemma~\ref{threshold}.
\footnote{For $n \geq 4B \geq 16$, hence $\log(n/B) \geq \log_B n$, else $n/B$ is a constant and the query can be answered in $O(1)$ I/O's.}

\subsection{Converting Top-{\large \lowercase{$\boldsymbol{{k}}$}}  to Threshold \\ via Logarithmic Sketch} \label{conversion}

We shall assume all scores are distinct and are within $[1, O(n)]$. Otherwise,  the ties can be broken arbitrarily and reduce the values into rank-space.

\paragraph{Marked nodes and Prime Nodes  in GST} 
We identify certain nodes in the $GST$ as marked nodes and prime nodes with respect to a parameter $g$ called the \emph{grouping factor}. The procedure starts by combining every $g$ consecutive leaves (from left to right) together as a group, and marking the lowest common ancestor (LCA) of first and last leaf in each group. Further, we mark the LCA of all pairs of marked nodes recursively. Additionally, we ensure that the root is always marked. At the end of this procedure, the number of marked nodes in $GST$ will be $O(n/g)$~\cite{HSV09}. 
Prime nodes are those which are the children of marked nodes. 
For any marked node $u^*$, there is a \emph{unique prime ancestor} node $u'$. In case $u^*$'s parent is marked then $u'=u^*$.
For every prime node $u'$, the corresponding marked descendant $u^*$ (if it exists) is unique. If $u'$ is marked then the descendant $u^*$ is same as $u'$.

Hon et. al.~\cite{HSV09} showed that, given any node $u$ with $u^*$ being its highest marked descendent (if exists), the number of leaves in the subtree of $u$, but not in the subtree of $u^*$ (which we call as fringe leaves) is at most $2g$. 
This means for a given threshold $\tau$, if $z$ is the number of outputs corresponding to $u^*$ as the locus node, then the number of outputs corresponding to $u$ as the locus is within $z\pm 2g$. 
This is because due to the additions leaves (possible each corresponds to a separate document), the score of at most $2g$ documents  can change (either decrease or increase). Therefore, we maintain the following information at every marked node $u^*$: the score of $q-$th highest scored link stabbed by $u^*$ for $q = 1,2,4,8,...$. The total space consumption is $O((n/g)\log n)=O(n)$ words by choosing $g=\log n$. 

Using the above values, the threshold $\tau$ corresponding to any given $u$ and $k$ can be computed as follows: firstly, find the highest marked node $u^*$ in the subtree of $u$ ($u^*=u$ if $u$ is marked). Now identify $i$ such that $2^{i-1}< k+2g= k+2\log n \leq 2^i$ and  choose $\tau$ as the score of $2^i$-th highest scored link stabbed by $u^*$ (which is pre-computed and stored). 
 This ensures that the number of outputs $z$ is at least $k$ and is at most $2k+O(g) = 2k+O(\log n)$. 
Note that the top-$k$ to threshold conversion can be performed in constant time and query I/O's in lemma~ will be $O(\log(n/B)+z/B)=O(\log (n/B)+(2k+\log n)/B) =O(\log(n/B)+k/B)$. 
From the retrieved $z$ outputs, the actual top-$k$ answers can be computed by selection and filtering in another $O(z/B)$ I/O's. 
We summarize our result in the following lemma. 

\BL
\label{lem_largek}
There exist a $O(n)$ word data structure for answering the following query in $O(\log(n/B)+k/B)$ I/O's: given a query point $u_P$ and an integer $k$, report the top-$k$ links that originate in the sub-tree of $u_P$ and target at a proper ancestor of $u_P$. \qed
\EL

It is easy to see that I/O bound in Lemma~\ref{lem_largek} is optimal for $k \geq B\log (n/B)$.
We derive special  structures for handling the case when $k <
B\log (n/B)$ in the next subsection.

\subsection{Special Structures for Bounded {\large \lowercase{$\boldsymbol{{k}}$}} } \label{bounded}

In this section, we derive a special structure for handling the case when $k$ is upper bounded by a parameter $g$.  
We summarize our result in the following lemma.

\BL
\label{lem_smallk}
There exists a $O(n)$ word data structure for answering top-$k$ queries for $k \leq g$ in $O(\log (g/B)+k/B)$ I/O's. 
\EL

Recall the definitions of marked nodes and prime nodes from Section~\ref{conversion}.
Let $u'$ be a prime node and $u^*$ (if it exists) be the unique highest marked descendent of $u'$ by choosing a grouping factor $g$ (which will be fixed later). 
All the links originating from the subtree of $u'$ are categorized into the following (see Figure 3).

\begin{itemize}
\item \emph{fringe-links:} The links originating from the subtree of $u'$, but not from the subtree of $u^*$. 

\item \emph{near-links:}  The links originating from the subtree of $u^*$ whose target is within the subtree of $u'$.

\item \emph{far-link:}  The link originating from the subtree of $u^*$ whose target is a proper ancestor of $u'$.

\item \emph{small-link:} The links with both origin and target within the subtree of $u^*$. 

\end{itemize} 

 \begin{figure}
\begin{center}
\includegraphics[scale=0.44]{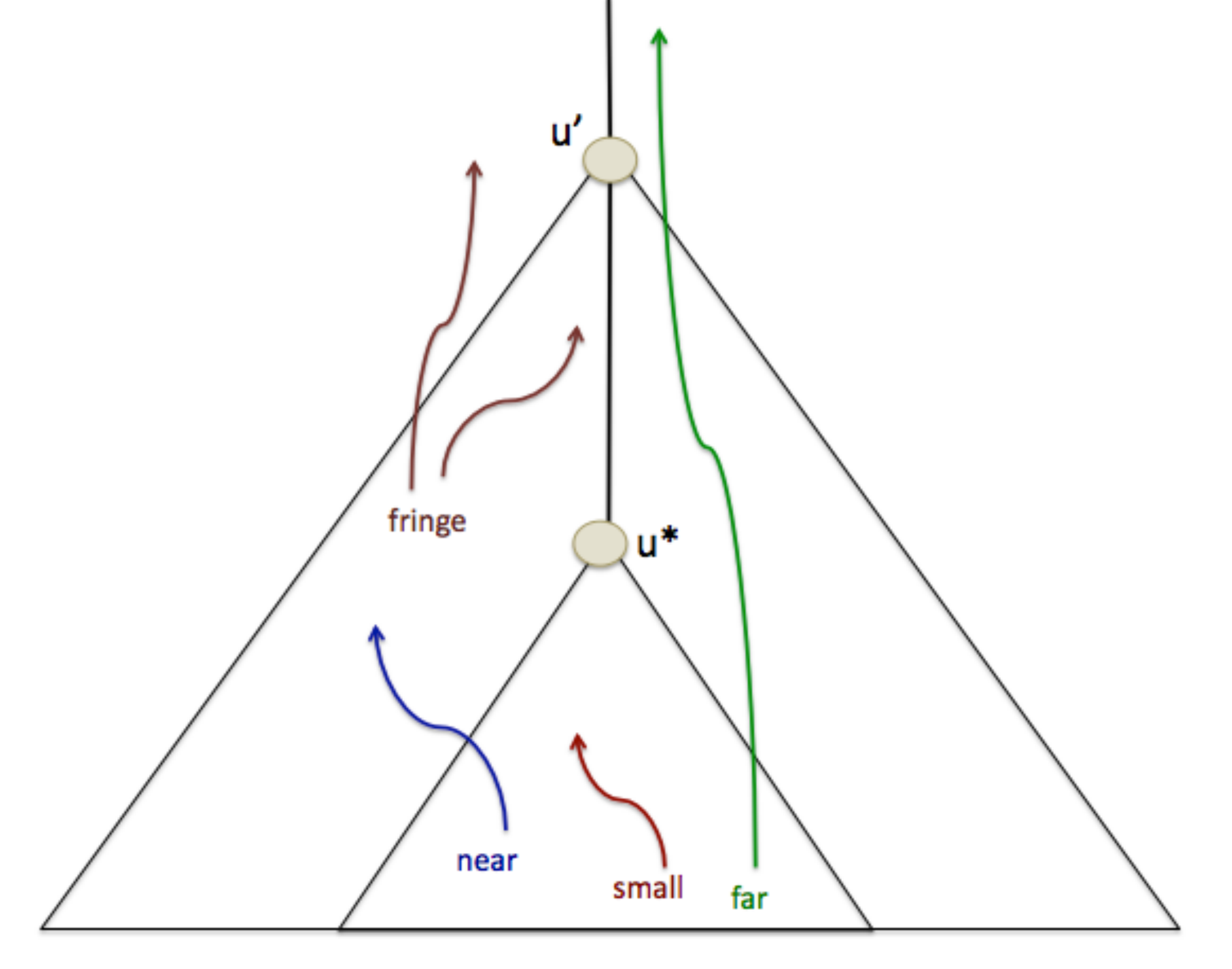}
\end{center}
\caption{Categorization of Links}
\end{figure}

\BL
\label{lem_countcand}
The number of \emph{fringe-links} and the number of \emph{near-links} of any prime node $u'$ is $O(g)$. 
\EL

\begin{proof} The number of leaves in $subtree(u')\backslash subtree(u^*)$ is at most $2g$~\cite{HSV09}. Only one link for each document comes out of the $subtree(u^*)$. Therefore, the number of \emph{fringe-links} can be bounded by $4g$. For every document $d$ whose link originates from $subtree(u^*)$ going out of it ends up as a \emph{near-link} if and only if $d$ exists at one of the leaves of $subtree(u')\backslash subtree(u^*)$. Thus, this can be bounded by $4g$ too. In the case that  $u^*$ does not exist for $u'$ only fringe-links exist and since the subtree size of $u'$ is $O(g)$ there can be no more than $O(g)$ of these links. \qed
\end{proof}

Consider the following set, consisting of $O(g)$ links with respect to $u'$: all \emph{fringe-links}, \emph{near-links} and $g$ highest scored \emph{far-links}.
We maintain these \emph{links} at $u'$ (as a data structure to be explained later). 
For any node $u$, whose closest prime ancestor (including itself) is $u'$, the above mentioned set is called \emph{candidate links} of $u$. From each $u$, we maintain the pointer to its closest prime ancestor where the set of \emph{candidate links} is stored.

\BL  
\label{lem_cl}
The \emph{candidate links} of any node $u$ contains top-$g$ highest scored links stabbed by $u$. 
\EL

\begin{proof}
Let $u'$ be the closest prime ancestor of $u$. 
If no marked descendant of $u'$ exist, then all the links are stored as candidate links.
Otherwise, \emph{small-links} can not ever be candidates as they never cross $u$.
Now, if $u$  lies on the path from $u'$ to $u^*$ then all \emph{far-links} will satisfy both origin and target conditions. Else, \emph{far-links} do not qualify. 
Hence, any link which is not among top-$g$ (highest scored) of these far-links, can never be the candidate. \qed
\end{proof} 

Taking a clue from Lemma~\ref{lem_countcand} and~\ref{lem_cl}, for every prime node $u'$, we shall maintain a data structure as in Lemma~\ref{lem_largek} by considering only the links stored at $u'$. Therefore, top-$k$ queries for any node $u$ with $u'$ being its lowest prime ancestor can be answered in $O(\log(g/B)+k/B)$ I/O's provided $k \leq g$.

\paragraph{Candidate Tree}
We define a candidate tree $CT(u')$ of node $u'$ (except the root) to be a modified version of subtree of $u'$ in GST augmented with candidate links stored at $u'$. Firstly, for every candidate link which is targeted above $u'$, we change the target to $v$, which will be a dummy parent of $u'$ in $CT(u')$. Now $CT(u')$ consists of those nodes which are either origin or target (after modification) of some candidate link of $u'$. Moreover, all the nodes in $subtree(u') \backslash subtree(u^*)$ are included as well. Since only the subset of nodes is selected from $subtree(u')$, our tree is basically a Steiner tree connecting these nodes. Moreover, the tree is edge-compacted so that no degree-1 node remains. Thus, the 
size of the tree as well as the number of associated links is $O(g)$.
Next we do a  rank-space reduction of  pre-order rank (w.r.t to GST) of  the nodes in $CT(u')$ as well as  the scores of candidate links.

The candidate tree (no degree-1 nodes) as well as the associated candidate links satisfies all the properties which we have exploited while deriving the  structure in Lemma~\ref{lem_largek}. 
Hence such a structure for $CT(u')$ can be maintained in $O(min(g, size(u'))$ words space and the top-$k$ links in $CT(u')$ stabbed by any node $u$, with $u'$ being its lowest prime ancestor can be retrieved in $O(\log(g/B)+k/B)$ I/O's. 
The total space consumption of structures corresponding all prime nodes can be bounded by $O(n)$ words as follows: 
 the number of prime nodes with at least a marked node in its subtree is $O(n/g)$, as each such prime node can be associated with a unique marked node. Thus the associated structures takes $O(n/g\times g)=O(n)$ words space. The candidate set of a prime node $u'$ with no marked nodes in its subtree consists of $O(size(u'))$ links, moreover a link cannot be in the candidate set of two such prime nodes. Thus the total space is $O(n)$ words in this case as well. This completes the proof of Lemma~\ref{lem_smallk}.

\subsection{Bootstrapping}
Optimal I/O bound of $O(1+k/B)$ can be achieved by using multiple structures as in Lemma~\ref{lem_smallk}. 
Clearly the structure in Lemma~\ref{lem_largek} is optimal for $k \geq B\log(n/B)$. However, for handling the case when $k < B\log(n/B)$, we shall use the structure in Lemma~\ref{lem_smallk} by choosing $g = B\log(n/B)$ and the query I/Os will be $O(\log(B\log(n/B)/B)+k/B)=O(\log\log(n/B)+k/B)$, which  is optimal for $k \geq B\log\log(n/B)$. For $k < B\log\log(n/B)$, we maintain another structure in Lemma~\ref{lem_smallk} with a different grouping factor $g=B\log\log(n/B)$ and obtain $O(\log\log\log(n/B)+k/B)$ query I/Os. 
In general, along with the structure in Lemma~\ref{lem_largek}, we maintain $h\leq \log^* n$ linear space structures in Lemma~\ref{lem_smallk} by choosing $g = B\log^{(b)}(n/B)$ for $b=1,2,3,...,h$. Thus any top-$k$ query with $k\geq B\log^{(h+1)}(n/B)$ can be answered optimally by querying on the structure corresponding to $g = B\log^{(b)}(n/B)$ , 
where $\log^{(b+1)}(n/B) < k/B \leq \log^{(b)}(n/B)$.
And if $k < B\log^{(h+1)}(n/B)$, we need additional $O(\log^{(h+1)}(n/B))$ I/O's. 
If we choose $h=\log^* n$, then $g = B$ and we need not store any
structure on $CT(u')$. Such a candidate tree fits entirely in constant number of blocks
which can be processed in $O(1)$ I/Os. This completes the proof of Theorem~\ref{the_result01}. 

\section{Internal Memory Structures}

This section shows how to solve top-$k$ document retrieval problem in word RAM model. 
The following theorem summarizes our result. 
\BT \label{finalll}
There exist a $O(n)$ word space data structure in word RAM model for solving top-$k$ document retrieval problem in $O(k)$, with documents retrieved in the decreasing order of the relevance. 
\ET

Firstly, we derive a linear space data structure with $O(\log n+k)$ query time, which is optimal for $k \geq \log n$, and the case when $k < \log n$ is handled separately using another data structure. 

\subsection{Structure for {\large \lowercase{$\boldsymbol{{k \geq \log n}}$}}}
For this, we shall choose $B=1$ and replace the substructures in our external memory structure by their internal memory counterparts. 
Top-$k$ to threshold conversion can be performed in constant time, whereas the number of outputs will be $O(k+\log n)$. 

\paragraph{Online sorted range reporting}
In the online sorted range reporting problem, an array $A$ is indexed so that given a query $(i,j)$, the entries in the subarray $A[i...j]$ can be reported in sorted order one by one until the user terminates the reporting. Brodal et.~al.~\cite{BFGL09} proposed a linear-space structure that achieves $O(1)$ cost per entry. Thus, by replacing the 3-sided range query structure by this structure, answers can be retrieved in the decreasing order of scores. 

\paragraph{Interval stabbing with priority}We shall use the following structure for Problem 2, which is capable of retrieving the answers in the decreasing order of score. 
\BL
Given a set $\cal{I}$ of weighted intervals, there exists a linear space data structure for answering the Interval stabbing with priority queries in $O(\log\log |\cal{I}|$$+k)$ time, where $k$ is the output size.
\EL

\begin{proof}

Let $\cal{I}$ be the set of weighted intervals. Consider a sweeping line that continuously moves from $-\infty$ to $+\infty$, on which a single-linked list is maintained to keep track of all the intervals that currently intersect the sweeping line. The intervals in the linked list are sorted in descending order of their weights. As the sweeping line encounters the left (resp., right) endpoint of an interval, it is inserted into (resp., deleted from) the linked list. For any stabbing query $u_P$, there must be a moment at which the first $k$ elements of the linked list are just the answer. To support query answering on all the snapshots, the linked list can be implemented with the persistent linked list~\cite{DSST89}. This structure guarantees that at any snapshot, once the list head has been identified, the linked list can be traversed in $O(1)$ time per element. Therefore, the top-$k$ intervals can be retrieved by first finding the list head of the correct snapshot, which is a predecessor search; then traversing the linked list at the snapshot. The space consumption is linear and the query time is $O(\log\log|\cal{I}| $$+k)$, where $O(\log\log |\cal{I}|)$ time is required to identify the list head in the persistent structure, and $O(k)$ time is taken to report the $k$ intervals.
\end{proof}

Thus, the top-$k$ document retrieval can be converted to at most $O(\log n)$ online sorted range reporting queries and one Interval stabbing with priority query. 
Note that each of this $O(\log n)$ subquery returns the answers in the sorted order. 
Therefore, the final (sorted) top-$k$ answers can be obtained by an $O(\log n)$-way merge. Since the number of elements in the heap for merging is $O(\log n)$, an atomic heap~\cite{FW94} can do each heap operation in $O(1)$ time in word RAM, leading to an overall $O(\log n+k)$ retrieval time. The space consumption is linear as all the sub structures used are of linear space.
\BL \label{ram-high}
There exists a linear space data structure in RAM model for solving top-$k$ document retrieval problem in $O(\log n+k)$ time.\qed
 \EL

\subsection{Structure for {\large \lowercase{$\boldsymbol{{k< \log n}}$}}}
The query time in Lemma~\ref{ram-high} is $O(k)$ for $k \geq \log n$. Therefore, queries with bounded $k$ (i.e., $k <\log n$) can be handled using another structure 
based on the framework in Section~\ref{bounded}. 
With $g=\log n$ as the grouping factor, we first identify the marked nodes and prime nodes in GST. We also construct the candidate tree along with the candidate links for each prime node. 
Our data structure for a particular candidate tree $CT(u')$ is simply the list of all associated candidate links in the decreasing order of its score, which we call as \emph{candidate list}.
We associate a bit vector $B_u$ of length $O(g) =O(\log n)$ with every node $u \in CT(u')$, such that $B_u[i]=1$ if and only if the $i$th highest scored link in the candidate list is stabbed by $u$. 
Constant time rank/select structures~\cite{RRS07} are also augmented with each $B_u$. 
In contrast to the external memory structure, the only difference here is, with each node in $GST$, we have a bit vector of space $O(\log n)$ bits, hence total $O(n\log n)$ bits or $O(n)$ words. 

In order to answer the top-$k$ query (for $k <g$), 
 we just retrieve those $select(B_{u_P}, i)$th highest scored links in the candidate list of $u_P$  for $i =1,2,3,...,k$, where $select(B_{u_P}, i)$ returns the position of the $i$th $1$ in the bit vector $B_{u_P}$. These select queries give the positions corresponding to the location of all  links stabbed by $u_P$. Since the links are sorted in the score order, we automatically get the top-$k$ answers in sorted order.

\BL \label{lem:small}
There exists a data structure taking $O(n)$ space which can answer top-$k$ document retrieval queries in optimal $O(k)$ time, for any $k < \log n$.
\qed
\EL

Combining Lemma \ref{ram-high} with Lemma \ref{lem:small} we obtain Theorem~\ref{finalll}.

\section{Conclusions}
For the seemingly 4-sided query, we showed the external memory bounds which are almost close to the bound obtainable on 3-sided query problem.
It remains to see if the $\log^*n$ factor from the space term can be eliminated. Also, to derive such results in cache-oblivious model will be important in the context of desktop searching systems. Extensions to higher dimensional range searching will be considered in future work.

\end{document}